\documentclass[12pt]{amsart}
\usepackage{amscd}
\usepackage{amsfonts}
\usepackage{epsf}
\usepackage{epic}
\usepackage{eepic}
\usepackage{latexsym}

\begin{document}

\title[The Relativistic Piecewise Uniform String]
{Casimir Energy and Thermodynamic Properties of the Relativistic
Piecewise Uniform String}

\author{I. Brevik}
\address{Division of Applied Mechanics, Norwegian University of Science 
and Technology, N-7034 Trondheim, Norway \,\, {\em E-mail address:}
{\rm iver.h.brevik@mtf.ntnu.no}}
\author{A.A. Bytsenko}
\address{Departamento de Fisica, Universidade Estadual de Londrina,
Caixa Postal 6001, Londrina-Parana, Brazil; on leave from Sankt-Petersburg
State Technical University, Russia \,\, {\em E-mail address:} 
{\rm abyts@fisica.uel.br}}

\date{February, 2000}

\thanks{First author partially supported by a CNPq grant (Brazil), RFFI 
grant (Russia) No 98-02-18380-a, and by GRACENAS grant (Russia) No 6-18-1997.}

\maketitle

\begin{abstract}

The Casimir energy for the transverse oscillations of a piecewise uniform 
closed string is calculated. The great adaptibility of this string model 
with respect 
to various regularization methods is pointed out. We survey several 
regularization methods:
the cutoff method, the complex contour integration method, and the 
zeta-function method. The
most powerful method in the present case is the contour integration method. 
The Casimir
energy turns out to be negative, and more so the larger is the number of 
pieces in the
string. The thermodynamic free energy $F$ is calculated for a two-piece 
string in the limit
when the tension ratio $x=T_{I}/T_{II}$ approaches zero.

\end{abstract}

\vspace{0.3cm}

\section{Introduction}

In the standard theory of closed strings - whatever the string is taken to 
be in Minkowski space
or in superspace - one usually assumes that the string is {\em homogeneous}, 
i.e. that the tension $T$ is the same everywhere. The {\em composite} string 
model, in which the string is assumed to consist of two or more 
separately uniform pieces, is a variant of the conventional theory. The 
system is relativistic, in the sense that the velocity $v_s$  of transverse 
sound is in each of the pieces assumed to be equal to the velocity of 
light: $v_s=\sqrt{T/\rho}=c$.
Here $T$ and $\rho$ (the density) refer to the piece under consideration. 
At each junction between pieces of different material there are two boundary 
conditions: the transverse displacement $\psi = \psi(\sigma,\tau)$ itself, 
as well as the transverse force $T\partial \psi/\partial \sigma$, must be 
continuous. Using the wave equation
$(\frac{\partial^2}{\partial\sigma^2}-\frac{\partial^2}{\partial\tau^2})
\psi=0$,
one can calculate the eigenvalue spectrum and the Casimir energy of the string.

The composite string model was introduced in 1990 \cite{brevniels90}; the 
string was there assumed to consist of two pieces $L_I$ and $L_{II}$. The 
dispersion equation was derived, and the Casimir energy calculated for 
various integer values of the length ratio $s=L_{II}/L_I$. Later on, the 
composite string model has been generalized and studied from various points 
of view [2-10]; we may mention, for instance, that the recent paper of Lu 
and Huang [9] discusses the Casimir energy for a composite Green - Schwarz 
superstring.

Some reasons why the composite string model turns out to be an attractive 
model to study are the following. First, if one performs Casimir energy 
calculations, one finds that the system is remarkably easy to regularize: 
one has access to the cutoff method [1], the complex contour integration 
method [3-5, 7], or the Hurwitz $\zeta-$ function method [2, 4, 5, 7] 
( [8] contains a review of the various regularization methods). As a physical 
result of the Casimir energy calculations it is also worth noticing that the 
energy is in general nonpositive, and is more negative the larger the number 
of uniform pieces in the string is.

The composite string model may moreover serve as a useful two-dimensional 
field theoretical model in general. The hope is that such a model can help us 
to understand the issue of the energy of the vacuum state in two-dimensional 
quantum field theories, what is quite a compelling goal. As a peculiar 
application, perhaps can this particular string model even play a role in the 
theories of the early universe. The notable point is here that the string can 
in principle adjust its zero point energy: the energy always becomes 
diminished if the string divides itself into a larger number of pieces.

It is also to be noted that there are strong formal similarities between this 
kind of theory and the phenomenological electromagnetic theory in material 
media satisfying the condition $\varepsilon \mu =1$, $\varepsilon$ denoting 
the permittivity and $\mu$ the permeability of the medium [11]. Obviously, 
the basic reason why the two theories become so similar is that the 
relativistic invariance is satisfied in both cases.

\section{Two-piece string}

\subsection{Dispersion relation}

Let the two junction points, lying at $\sigma=0$ and $\sigma=L_I$, separate 
the type $I$ and type $II$ pieces from each other. The total length of the 
closed string is $L=L_I+L_{II}$. We define $x$ to be the tension ratio and 
define also the function $F(x)$:

$$
x=\frac{T_I}{T_{II}},~~~~~F(x)=\frac{4x}{(1-x)^2}
\mbox{.}
\eqno{(2.1)}
$$
The dispersion equation becomes

$$
F(x)\sin^2\left(\frac{\omega L}{2}\right)+\sin \omega L_I\sin \omega L_{II}=0
\mbox{.}
\eqno{(2.2)}
$$
The Casimir energy $E$ of the system is defined as the zero-point energy $E_{I+II}$ of the two parts, minus the zero-point energy of the uniform string:

$$
E=E_{I+II}-E_{\rm uniform}=\frac{1}{2}\sum \omega_n-E_{\rm uniform}
\mbox{.}
\eqno{(2.3)}
$$
Here the sum goes over all eigenstates, with account of their degeneracy. It 
is irrelevant whether $E_{\rm uniform}$ is calculated for type $I$ material 
or type {II} material in the string, the reason for this being the 
relativistic invariance. We will consider three different methods for 
regularizing the Casimir energy. 

\subsection{Cutoff regularization}

The simplest way to proceed [1] is to introduce a function $f=\exp 
(-\alpha \omega_n)$, with $\alpha$ a small positive parameter, and to 
multiply the nonregularized expression for $E$ by $f$ before summing over 
the modes.

We consider first the case of a {\it uniform} string, corresponding to $x=1$. 
The dispersion equation (2.2) yields the eigenvalue spectrum $\omega L=1$, 
which means

$$
\omega_n=2 \pi n/L,~~~~n=1,2,3,...
\eqno{(2.4)}
$$
Taking into account that these modes are degenerate, we find for the 
zero-point energy

$$
E_{\rm uniform}=\frac{L}{2\pi\alpha^2}-\frac{\pi}{6L}+{\mathcal O}(\alpha^2)
\mbox{.}
\eqno{(2.5)}
$$
Let us next consider the limiting case $x \rightarrow 0$ (we let $T_I\rightarrow 0$ while keeping $T_{II}$ finite). The dispersion relation allows two sequences of modes,

$$
\omega_n=\pi n/L_I,~~~~~~\omega_n=\pi n/L_{II},~~~~n=1,2,3,...
\eqno{(2.6)}
$$
If $s$ denotes the length ratio, $s=L_{II}/L_I$, we then get the simple 
formula for the Casimir energy

$$
E=-\frac{\pi}{24 L}(s+\frac{1}{s}-2)
\mbox{.}
\eqno{(2.7)}
$$
Now let $s$ be an {\it odd} integer. The dispersion equation yields one 
degenerate branch, determined by 

$$
\sin\omega L_I=0,~~~~\omega L_I=\pi n
\mbox{,}
\eqno{(2.8)}
$$
and there are in addition $\frac{1}{2}(s-1)$ nondegenerate double branches, 
determined by solving an algebraic equation of degree $\frac{1}{2}(s-1)$ in   
$\sin^2\omega L_I$. The frequency spectrum can be expressed as 

$$
\omega L_I=\left\{ \begin{array}{ll}
                    \pi (n+\beta),\\
                     \pi (n+1-\beta),
                     \end{array}
            \right.
\eqno{(2.9)}
$$
where $n=0,1,2,...$, and where $\beta$ is a number in the interval 
$0< \beta \leq \frac{1}{2}$. Each double branch yields the four solutions 
$\pi \beta$, $\pi (1-\beta)$, $\pi(1+\beta)$, and $\pi(2-\beta)$ for 
$\omega L_I$ in the region between $0$ and $2\pi$. 

Introducing for convenience the abbreviation $t=\pi\alpha(s+1)/L$, we obtain

$$
E({\rm degenerate~~ branch})=\frac{1}{\alpha t}-\frac{t}{12\alpha}+
{\mathcal O}(t^2),
\eqno{(2.10)}
$$
$$
E({\rm double~~ branch})=\frac{1}{\alpha t}+\frac{t}{6\alpha}-
\frac{t}{4\alpha}[\beta^2+(1-\beta)^2]+ {\mathcal O}(t^2).
\eqno{(2.11)}
$$
We replace $\beta$ by $\beta_i$, sum (2.11) over all 
$\frac{1}{2}(s-1)$ double branches, and add (2.10) to obtain $E_{I+II}$.
Subtracting off the uniform string result (2.5), and letting 
$t \rightarrow 0$, we get the Casimir energy for odd $s$, 

$$
E=\frac{\pi s(s-1)}{12L}-\frac{\pi (s+1)}{4L}\sum_{i=1}^{(s-1)/2}[\beta_i^2+(1-\beta_i)^2].
\eqno{(2.12)}
$$
The cutoff terms drop out.

If $s$ is an {\it even} integer, we obtain by an analogous argument

$$
E=\frac{\pi s(2s+1)}{6L}-\frac{\pi(s+1)}{8L}\sum_{i=1}^s
[\beta_i^2+(2-\beta_i)^2],
\eqno{(2.13)}
$$
where now each $\beta_i$ lies in the interval $0 < \beta_i \leq 1$.

\subsection{Contour integration method}

This is a very powerful method. In the context of Casimir calculations it 
dates back to van Kampen et al. [12]. The method was first applied to the 
composite string system in Ref. [3]. The starting point is the so-called
argument principle, which states that any meromorphic function  $g(\omega)$  
satisfies the relation

$$
\frac{1}{2\pi i}\oint \omega \frac{d}{d\omega}\ln g(\omega)=
\sum\omega_0-\sum\omega_{\infty},
\eqno{(2.14)}
$$
where $\omega_0$ are the zeros and $\omega_{\infty}$ are the poles of  
$g(\omega)$ inside the integration contour. The contour is chosen to be a 
semicircle of large radius $R$ in the right half complex $\omega$ plane, 
closed by a straight line from $\omega=iR$ to $\omega=-iR$. The great 
advantage of the method - in contradistinction to the previous cutoff method -
is that the {\it multiplicity} of the zeros (there are no poles in the 
present case) are automatically taken care of.

We make the following ansatz for $g(\omega)$:

$$
g(\omega)=\frac{F(x)\sin^2[(s+1)\omega L_I/2]+\sin(\omega L_I)
\sin(s\omega L_I)}{F(x)+1}.
\eqno{(2.15)}
$$
This means that $g(\omega)$ is chosen to be the expression to the left in 
(2.2), multiplied by $[F(x)+1]^{-1}$. This choice is convenient, since it 
allows us to perform partial integrations in the energy integral without 
encountering any divergences in the boundary terms when 
$R \rightarrow \infty$. The final result becomes $(\omega=i\xi)$

$$
E=\frac{1}{2\pi}\int_0^{\infty}\ln \left|\frac{F(x)+\frac{\sinh \xi L_I 
\sinh s\xi L_I}{\sinh^2[(s+1)\xi L_I/2]}}{F(x)+1} \right| d\xi.
\eqno{(2.16)}
$$
This zero-temperature result is very general; it holds for any value of $s$, 
not only for integers $s$ as considered in the previous subsection. Since 
(2.16) is invariant under the interchange $s \rightarrow 1/s$, it follows 
that $s$ can be restricted to the interval $s \geq 1$ without any loss of 
generality. If $x\rightarrow 0$, we recover the simple formula (2.7).

Another advantage of the contour integration method is that the 
zero-temperature result can easily be generalized to the case of finite 
temperatures. The integration over continuous imaginary frequencies 
$\xi$ then has to be replaced by a sum over discrete Matsubara frequencies 
$\xi_n=2\pi nk_BT,~~n=0, 1, 2,...$ We get

$$
E(T)=k_BT{\sum_{n=0}^{\infty}}'\ln \left|\frac{F(x)+\frac{\sinh \xi_nL_I\sinh s\xi_n L_I}
{\sinh^2[(s+1)\xi_nL_I/2]}}{F(x)+1} \right|,
\eqno{(2.17)}
$$ 
valid for any temperature $T$. The prime on the summation sign means that the $n=0$ term is taken with half weight.

\subsection{$\zeta-$ function method}

This elegant regularization method has proved to be most useful in many 
cases. General treatises on it can be found in Refs. [13, 14]. The first 
application to the composite string was made by Li et al. [2]. The 
appropriate $\zeta-$function to be used in this case is not the Riemann 
function $\zeta(s)$, but instead the Hurwitz function $\zeta(s,a)$, the 
latter being originally defined as

$$
\zeta(s,a)=\sum_{n=0}^{\infty}(n+a)^{-s}~~~~(0<a<1,~~~{\rm Re}~s>1).
\eqno{(2.18)}
$$
For practical purposes on needs only the property

$$
\zeta(-1,a)=-\frac{1}{2}\left(a^2-a+\frac{1}{6}\right)
\eqno{(2.19)}
$$
of the analytically continued Hurwitz function.

The $\zeta-$function method has one important property in common with the 
cutoff method: the eigenvalue spectrum must be determined explicitly. 
Consider the uniform string first: in this case the Riemann function is 
adequate, giving the zero-point energy

$$
E_{\rm uniform}=\frac{2\pi}{L}\zeta(-1)=-\frac{\pi}{6L},
\eqno{(2.20)}
$$
in agreement with the finite part of (2.5). Consider next the composite 
string, assuming $s$ to be an odd integer: by inserting the degenerate 
branch eigenvalue spectrum (2.8) we have

$$
E({\rm degenerate~~branch})=-\frac{\pi}{12L_I}.
\eqno{(2.21)}
$$
Using the generic form (2.9) for the double branches we obtain analogously

$$
E({\rm double~~branch}) = \frac{\pi}{2L_I}[\zeta(-1,\beta)+\zeta(-1,1-\beta)]
$$
$$
= \frac{\pi}{6L_I}-\frac{\pi}{4L_I}[\beta^2+(1-\beta)^2].
\eqno{(2.22)}
$$
Summing (2.22) over the $\frac{1}{2}(s-1)$ double branches, and adding (2.21), 
we obtain the composite string's zero-point energy

$$
E_{I+II}=\frac{\pi (s-2)}{12 L_I}-\frac{\pi}{4 L_I}\sum_{i=1}^{(s-1)/2}
[\beta_i^2+(1-\beta_i)^2].
\eqno{(2.23)}
$$
Now subtracting off (2.20), we obtain the same expression for the Casimir 
energy $E$ as in Eq. (2.12).

The case of even integers $s$ is treated analogously. The $\zeta-$function 
method is somewhat easier to implement than the cutoff method.

\subsection{$\zeta-$function regularization for some infinite products}

Let us consider a method of regularization for the infinite products of the
form:

$$
{\frak P}=\prod_{n=1}^{\infty}\left(\frac{n}{b}+a\right),
\eqno{(2.24)}
$$
$$
P=\prod_{n=1}^{\infty}\left(\frac{n^2}{B}+A\right)
=\prod_{n=1}^{\infty}\left(\frac{n}{\sqrt{B}}+i\sqrt{A}\right)
\left(\frac{n}{\sqrt{B}}-i\sqrt{A}\right)
\mbox{,}
\eqno{(2.25)}
$$
where $a, b$ are real numbers, $A, B>0$. The $\zeta-$function associated with
the product (2.24) has the form

$$
\zeta_{\frak P}(s)=\sum_{n=1}^{\infty}\left(\frac{n}{b}+a\right)^{-s}
=\frac{1}{\Gamma(s)}\sum_{n=1}^{\infty}\int_0^{\infty}t^{s-1}
e^{-t\left(\frac{n}{b}+a\right)}dt
$$
$$
=\frac{1}{\Gamma(s)}\int_0^{\infty}t^{s-1}
\frac{e^{-ta}}{1-e^{-t/b}}dt-a^{-s}
\mbox{.}
\eqno{(2.26)}
$$
Then the following equation holds

$$
\zeta_{\frak P}(s)=b^s\zeta(s,ab)-a^{-s}
\mbox{.}
\eqno{(2.27)}
$$
Using the equations

$$
\zeta(0,\ell)=\frac{1}{2}-\ell
\mbox{,}
\eqno{(2.28)}
$$
$$
\frac{d}{ds}\zeta(s,\ell)|_{s=0}=\log \Gamma(\ell)-\frac{1}{2}\log 2\pi
\mbox{,}
\eqno{(2.29)}
$$
and Eq. (2.27) we have

$$
\frac{d}{ds}\zeta_{\frak P}(s,\ell)|_{s=0}=\zeta(0,ab)\log b
+\frac{d}{ds}\zeta(0,ab)+\log a
$$
$$
=(\frac{1}{2}-ab)\log b +\log \Gamma(ab)+ \log a -\frac{1}{2}\log 2\pi
\mbox{.}
\eqno{(2.30)}
$$
Therefore 

$$
{\frak P}=\exp\left\{\sum_{n=1}^{\infty}\log\left(\frac{n}{b}+a\right)\right\}
=\exp\left\{-\frac{d}{ds}\zeta_{\frak P}(0)\right\}
=\frac{\sqrt{2\pi}}{a\Gamma(ab)}b^{ab-\frac{1}{2}}
\mbox{,}
\eqno{(2.31)}
$$
and finally

$$
P=\frac{2\pi}{A\sqrt{B}\,\Gamma(i\sqrt{AB})\Gamma(-i\sqrt{AB})}
=\frac{2}{\sqrt{A}}\sinh (\pi\sqrt{AB})
\mbox{.}
\eqno{(2.32)}
$$

Let $\eta(\tau)$ is the Dedekind $\eta$-function, 
$$
\eta(\tau) = e^{\frac{\pi i\tau}{12}}\prod_{n=1}^{\infty}(1 - e^{2\pi in\tau})
\mbox{,}
\eqno{(2.32)}
$$
$$
\eta(i\tau)=\frac{1}{\sqrt{2}}\prod_{n=1}^{\infty}\sinh (\pi n\tau)
\mbox{.}
\eqno{(2.33)}
$$
Then we can perform the computations in two different orders:

$$
\prod_{n,m=1}^{\infty}\left(\frac{m^2}{a^2}+\frac{n^2}{b^2}\right)=
\prod_{m=1}^{\infty}\frac{2a}{m}\sinh\left(\pi\frac{mb}{a}\right)
=\prod_{n=1}^{\infty}\frac{2b}{n}\sinh\left(\pi\frac{na}{b}\right)
\mbox{,}
\eqno{(2.34)}
$$
which implies very well-known modular property of the eta function:
$\sqrt{b}\eta(ib/a)=\sqrt{a}\eta(ia/b)$. 

By analogy with Eq. (2.24) we can consider
${\mathcal P}=\prod_{n=1}^{\infty}\left[(2n+1)/b+a\right]$. The following
formula holds:

$$
\prod_{n=0}^{\infty}\left[\frac{(2n+1)^2}{B}+A\right]
=\prod_{n=0}^{\infty}\left[\frac{2n+1}{\sqrt{B}}+i\sqrt{A}\right]
\left[\frac{2n+1}{\sqrt{B}}-i\sqrt{A}\right]
$$
$$
=2\cosh \left(\frac{\pi\sqrt{AB}}{2}\right)
\mbox{.}
\eqno{(2.35)}
$$

\section{$2N-$piece string}

\subsection{Recursion equation and casimir energy}

In the same way one can consider the Casimir theory of a string of length 
$L$ divided into three pieces, all of the same length.  The theory for this 
case has been given in Refs. [5] and [8]. Here, we shall consider instead a 
string divided into $2N$ pieces of equal length, of alternating type $I$/type 
$II$ material. The string is relativistic, in the same sense as before. The 
basic formalism for arbitrary integers $N$ was set up in Ref. [4], but the 
Casimir energy was there calculated in full only for the case of $N=2$. A 
full calculation was worked out in Ref. [7]; cf. also Ref. [8]. A key point 
in [7] was the derivation of a new recursion formula, which is applicable 
for general integers $N$.

We introduce two new symbols, $p_N$ and $\alpha$:

$$
p_N=\omega L/N,~~~~~\alpha=(1-x)/(1+x).
\eqno{(3.1)}
$$
The eigenfrequencies are determined from

$$
{\rm Det}[{\bf M}_{2N}(x,p_N)-{\bf 1}]=0.
\eqno{(3.2)}
$$
Here it is convenient to scale the resultant matrix $ {\bf M}_{2N}$ as

$$
{\mathbf M}_{2N}(x,p_N)=\left[ \frac{(1+x)^2}{4x}\right]^N{\mathbf m}_{2N}
(\alpha, p_N),
\eqno{(3.3)}
$$
and to write $ {\bf m}_{2N} $ as a product of component matrices:

$$
{\bf m}_{2N}(\alpha, p_N)=\prod_{j=1}^{2N}{\bf m}^{(j)}(\alpha, p_N),
\eqno{(3.4)}
$$
with

$$
{\mathbf m}^{(j)}(\alpha,p_N)=\left( \begin{array}{ll}
                                   1, & \mp \alpha e^{-ijp_N}\\
                                   \mp \alpha e^{ijp_N}, & 1
                                     \end{array} \right)
\eqno{(3.5)}
$$
for $j=1, 2,...(2N-1)$. The sign convention is to use +/- for even/odd $j$. At the last junction, for $j=2N$, the component matrix has a particular form (given an extra prime for clarity):

$$
{\mathbf m}'_{2N}(\alpha,p_N)=\left( \begin{array}{ll}
                                   e^{-iN p_N}, & \alpha e^{-i N p_N}\\
                                   \alpha e^{iN p_N}, & e^{iN p_N}
                                    \end{array} \right) .
\eqno{(3.6)}
$$
Now the recursion formula alluded to above can be stated:

$$
{\mathbf m}_{2N}(\alpha, p_N)={\mathbf \Lambda}^N(\alpha, p_N),
\eqno{(3.7)}
$$
where $ \mathbf \Lambda $ is the matrix

$$
{\mathbf \Lambda}(\alpha,p)=\left( \begin{array}{ll}
                                 a & b\\
                                 b^* & a^*
                                 \end{array} \right) ,
\eqno{(3.8)}
$$
with

$$
a=e^{-ip}-\alpha^2, ~~~~~b=\alpha (e^{-ip}-1).
\eqno{(3.9)}
$$
The obvious way to proceed is now to calculate the eigenvalues of 
$\mathbf \Lambda$, and express the elements of $ {\mathbf M}_{2N}$ as powers 
of these. More details can be found in [7].

Consider next the Casimir energy. The most powerful regularization method, 
as above, is the contour regularization method. Using it we obtain, for 
arbitrary $x$ and arbitrary integers $N$, at zero temperature,

$$
E_N(x)=\frac{N}{2\pi L}\int_0^{\infty}\ln \left| \frac{2(1-\alpha^2)^N-
[\lambda_+^N(iq)+\lambda_-^N(iq)]}{4 \sinh^2 (Nq/2)} \right| dq.
\eqno{(3.10)}
$$
Here $\lambda_{\pm}$ are eigenvalues of $\mathbf \Lambda$, for imaginary 
arguments $iq$, of the dispersion equation. Explicitly,

$$
\lambda_{\pm}(iq)=\cosh q-\alpha^2 \pm [(\cosh q-\alpha^2)^2-(1-\alpha^2)^2]^
\frac{1}{2}.
\eqno{(3.11)}
$$
Evaluation of the integral shows that $E_N(x)$ is negative, and the more so 
the larger is $N$. A string can thus in principle always diminish its 
zero-point energy by dividing itself into a larger number of pieces of 
alternating type I/II material.

In the limiting case of $x \rightarrow 0$ the integral can be solved exactly:

$$
E_N(0)=-\frac{\pi}{6 L}(N^2-1).
\eqno{(3.12)}
$$
The generalization of (3.10) to the case of finite temperatures is easily 
achieved following the same method as above. 

As an alternative method, on can instead of contour integration make use of 
the $\zeta-$ function method; one then has to determine the spectrum 
explicitly and thereafter put in the degeneracies by hand. The latter mehod 
is therefore most suitable for low $N$.

\subsection{Scaling invariance}

A rather unexpected scaling invariance property of the Casimir energy becomes 
apparent if we examine the behaviour of the function $f_N(x)$ defined by

$$
f_N(x)=\frac{E_N(x)}{E_N(0)}.
\eqno{(3.13)}
$$
This function generally has a value that lies between zero and one. If we 
calculate $E_N(x)$ (usually numerically) versus $x$  for some fixed value of 
$N$, we find that the resulting curve for  $f_N(x)$ is practically the 
{\em same}, irrespective of the value of $N$, as long as $N \geq 2$. (The 
case $N=1 $ is exceptional, since $E_1(x)=0$.) Numerical trials show that the 
simple analytical form 

$$
f_N(x) \rightarrow f(x)= (1-\sqrt x )^{5/2}
\eqno{(3.14)}
$$
is a useful approximation, in particular in the region $0<x<0.45$.

\section{Planar oscillations of the classical string in the minkowski
space}

We begin by considering the classical theory of the oscillating two-piece
string in the Minkowski space. The total length of the string is $L$.
For later purpose we shall set  $L = \pi$.  With $L_I$, $L_{II}$ denoting the 
length of the two pieces, we
thus have $L_I + L_{II} = \pi$. As mentioned the string is relativistic,
in the sense that
the velocity $v_s$ of transverse sound is everywhere required to be equal
to the velocity of light ($\hbar = c = 1$): $v_s = (T_I/\rho_I)^{1/2} = 
(T_{II}/\rho_{II})^{1/2} = 1$.
Here $T_I, T_{II}$ are the tensions and $\rho_I, \rho_{II}$ are the mass
densities of the two pieces. We let $s$ denote the length ratio and $x$ the
tension ratio: $s = L_{II}/L_I,\,\,\,\,\,\,\, x = T_I/T_{II}$.
Assume now that the transverse oscillations of the string, called
$\psi(\sigma,\tau)$, are linear, and take place in the plane of the string.
(We employ usual notation, so that $\sigma$ is the position coordinate and
$\tau$ the time coordinate of the string.) We can thus write in the two
regions

$$
\psi_I = \xi_I e^{i\omega(\sigma-\tau)} + \eta_I e^{-i\omega(\sigma+\tau)}
\mbox{,}
\eqno{(4.1)}
$$
$$
\psi_{II} = \xi_{II} e^{i\omega(\sigma-\tau)} + \eta_{II} e^{-i\omega(\sigma+\tau)}
\mbox{,}
\eqno{(4.2)}
$$
with the $\xi$ and $\eta$ being constants. Taking into account the junction
conditions at $\sigma = 0$ and $\sigma = L_I$, meaning that $\psi$ itself as 
well as the transverse force $T \partial \psi / \partial \sigma$ be 
continuous, we obtain the dispersion equation

$$
\frac{4x}{(1-x)^2} \sin^2\frac{\omega\pi}{2} +
\sin \left( \frac{\omega\pi}{1+s} \right)
\sin \left( \frac{\omega s \pi}{1+s} \right) = 0
\mbox{.}
\eqno{(4.3)}
$$
From this equation the eigenvalue
spectrum can be calculated, for arbitrary values of $x$ and $s$. Because of
the invariance under the substitution $x \rightarrow 1/x$, one can restrict
the ratio $x$ to lie in the interval $0 < x \leq 1$. Similarly, because
of the invariance under the interchange $L_I \leftrightarrow L_{II}$ one
can take $L_{II}$ to be the larger of the two pieces, so that
$s \geq 1$.

In the following we shall impose two simplifying conditions: (i) We take the
tension ratio limit to approach zero, $x \rightarrow 0$.
Assuming $T_{II}$ to be a finite quantity, this limit implies that
$T_I \rightarrow 0$.  From the junction conditions given in 
\cite{brevniels90} we obtain in this limit the equations

$$
\xi_I + \eta_I = \xi_{II} e^{i\pi\omega} + \eta_{II} e^{-i\pi\omega}
\mbox{,}
\eqno{(4.4)}
$$

$$
\xi_I e^{2\pi i \omega /(1+s)} + \eta_I = \xi_{II} e^{2\pi i \omega /(1+s)} 
+ \eta_{II}
\mbox{,}
\eqno{(4.5)}
$$

$$
\xi_{II} e^{2\pi i \omega} = \eta_{II}
\mbox{,}
\eqno{(4.6)}
$$

$$
\xi_{II} e^{2\pi i \omega /(1+s)} = \eta_{II}
\mbox{.}
\eqno{(4.7)}
$$
According to the dispersion equation (4.3) we obtain now two sequences
of modes.
The eigenfrequencies are seen to be proportional to integers $n$, and will
for clarity be distinguished by separate symbols $\omega_n(s)$ and
$\omega_n(s^{-1})$:

$$
\omega_n(s) = (1+s)n
\mbox{,}
\eqno{(4.8)}
$$

$$
\omega_n(s^{-1}) = (1+ s^{-1})n
\mbox{,}
\eqno{(4.9)}
$$
with $n = \pm 1, \pm 2, \pm 3,...$, corresponding to the first and
the second branch.

(ii) Our second condition is that the length ratio $s$ is an integer,
$s = 1,2,3,\cdots$.

\section{Classical string in flat $D$-dimensional spacetime}

\subsection{Oscillator coordinates. The hamiltonian}

We are now able to generalize the theory. We consider henceforth the motion
of a two-piece classical string in flat $D$-dimensional space-time.
Following the notation in \cite{green87}
we let $X^\mu(\sigma,\tau)\,\,\,(\mu = 0,1,2,\cdots (D-1))$ specify the 
coordinates on the world sheet. For each of the two
branches - corresponding to Eqs. (4.8) and (4.9) respectively -
we can write the general expression for $X^\mu$ in the form

$$
X^\mu = x^\mu + \frac{p^\mu\tau}{\pi \bar{T}(s)} + \theta(L_I - \sigma)
X_I^\mu + \theta(\sigma - L_I)X_{II}^\mu
\mbox{,}
\eqno{(5.1)}
$$
where $x^\mu$ is the center of mass position and $p^\mu$ is the total
momentum of the string. Besides $\bar{T}(s)$ denotes the mean tension,

$$
\bar{T}(s) = \frac{1}{\pi}(L_IT_I + L_{II}T_{II})~ \rightarrow~
\frac{s}{1+s}T_{II}
\mbox{.}
\eqno{(5.2)}
$$
The second term in (5.1) implies that the string's translational
energy $p^0$ is set equal to $\pi \bar{T}(s)$. This generalizes the relation
$p^0 = \pi T$ that is known to hold for a uniform string \cite{green87}.
The two last terms in (5.1) contain the step function,
$\theta(x > 0) = 1,\,\,\,\theta(x < 0) = 0$. To show the structure of the 
decomposition
of $X^\mu$ into fundamental model we give here the expressions for $X_I^\mu$
for each of the two branches: for the first branch

$$
X_I^\mu = \frac{i}{2} l(s) \sum_{n \neq 0} \frac{1}{n} \left[
\alpha_n^\mu(s) e^{i(1+s)n(\sigma-\tau)} + \tilde{\alpha}_n^\mu(s)
e^{-i(1+s)n(\sigma+\tau)}\right]
\mbox{,}
\eqno{(5.3)}
$$
where the $\alpha_n, \tilde{\alpha}_n$ are oscillator coordinates of the 
right- and left-moving waves respectively. The sum over $n$ goes over all 
positive and
negative integers except from zero. The factor $l(s)$ is a constant. For the
second branch in region I, analogously

$$
X_I^\mu=\frac{i}{2} l(s^{-1})\sum_{n \neq 0} \frac{1}{n} \left[
\alpha_n^\mu(s^{-1}) e^{i(1+s^{-1})n(\sigma-\tau)}
+ \tilde{\alpha}_n^\mu(s^{-1})e^{-i(1+s^{-1})n(\sigma+\tau)} \right]
\mbox{,}
\eqno{(5.4)}
$$
where $l(s^{-1})$ is another constant, which in principle can be different 
from $l(s)$. Since $X^\mu$ is real, we must have

$$
\alpha_{-n}^\mu = (\alpha_n^\mu)^*, ~~~~\tilde{\alpha}_{-n}^\mu =
(\tilde{\alpha}_n^{\mu})^*
\mbox{.}
\eqno{(5.5)}
$$
When writing expressions (5.3) and (5.4), we made use of Eqs.
(4.8) and (4.9) for the eigenfrequencies. The condition $x \rightarrow 0$
was thus used. The condition that $s$ be an integer has however not so
far been used. This condition will be of importance when we construct
the expression for $X_{II}^\mu$. Before doing this, let us however
consider the constraint equation for the composite string.
Conventionally, when the string is uniform the two-dimensional
energy-momentum tensor $T_{\alpha \beta}\,\,\, (\alpha,\beta = 0,1)$, 
obtainable as
the variational derivative of the action $S$ with respect to the
two-dimensional metric, is equal to zero. In particular, the
energy density component is then $T_{00} = 0$ locally. In the
present case the situation is more complicated, due to the fact
that the presence of the junctions restricts the freedom of the
variations $\delta X^\mu$. We cannot put $T_{\alpha \beta} = 0$ locally
anymore. What we have at our disposal, is the expression for the
action

$$
S = -\frac{1}{2}\int d\tau d\sigma T(\sigma) \eta^{\alpha \beta} 
\partial_\alpha X^\mu\partial_\beta X_\mu
\mbox{,}
\eqno{(5.6)}
$$
where $T(\sigma)$ is the position-dependent tension

$$
T(\sigma) = T_I + (T_{II} - T_I) \theta(\sigma - L_I)
\mbox{.}
\eqno{(5.7)}
$$
The momentum conjugate to $X^\mu$ is $P^\mu(\sigma) = T(\sigma) \dot{X}^{\mu}$.
The Hamiltonian of the two-dimensional sheet becomes accordingly
(here $L$ is the Lagrangian)

$$
H = \int_{0}^{\pi} \left[ P_\mu(\sigma) \dot{X}^\mu - L \right] d\sigma =
\frac{1}{2} \int_{0}^{\pi} T(\sigma) (\dot{X}^2 + {X'}^2)d\sigma
\mbox{.}
\eqno{(5.8)}
$$
The basic condition that we shall impose, is that $H = 0$ when
applied to the physical states. This is a more weak condition than
the strong condition $T_{\alpha \beta} = 0$ applicable for a uniform string.

\subsection{Classical mass formula. The first branch}

Assume  that $s$ is an arbitrary integer,  $s = 1,2,3,\cdots$. When $s$ is 
different from 1, we have to
distinguish between the eigenfrequencies $\omega_n(s)$ and 
$\omega_n(s^{-1})$ for the first and the second branch.
Let us consider the first  branch. In region I, the representation for the 
right- and left-moving modes was given above, in Eq. (5.3). For reasons 
that will become clear from the quantum mechanical discussion later, we 
will choose $l(s)$ equal to $l(s) = (\pi T_I)^{-1/2}$.
Since we have assumed $T_I$ to be small, that expression will tend to 
infinity. 

When writing the analogous mode expansion in region II, we have to observe the
junction conditions (4.4) - (4.7), which hold for all $s$. For the 
first branch $\omega_n(s)$, and for {\it odd} values of $s$, it is seen that 
the junction conditions impose no
restriction on the values of $n$. {\it All} frequencies, corresponding to
$n = \pm1,\pm2,\pm3,\cdots$, permit the waves to propagate
from region I to region II. Equations (4.4) - (4.7) reduce in
this case to the equations

$$
\xi_{I} + \eta_{I} = 2\xi_{II} = 2\eta_{II}
\mbox{,}
\eqno{(5.9)}
$$
which show that the right- and left-moving amplitudes $\xi_{I}$ and
$\eta_{I}$ in region I can be chosen freely and that the amplitudes
$\xi_{II},\eta_{II}$ in region II are thereafter fixed. Transformed into
oscillator coordinate language, this means that $\alpha_n^\mu$ and
$\tilde{\alpha}_n^\mu$ can be chosen freely.

If $s$ is an {\it even} integer, then the validity of Eqs. (5.9) requires $n$
in Eq. (4.8) to be even. If $n$ is odd, the junction conditions reduce instead
to

$$
\xi_I + \eta_I = 0, \,\,\,   \xi_{II} = \eta_{II} = 0
\mbox{,}
\eqno{(5.10)}
$$
which show that the waves are now unable to penetrate into region II. 
The oscillations in region I are in this case standing waves.

The expansion for the first branch in region II can in view of (5.9)
be written

$$
X_{II}^\mu = \frac{i}{2\sqrt{\pi T_I}} \sum_{n \neq 0} \frac{1}{n}
\gamma_n^\mu(s)e^{-i(1+s)n \tau} \cos[(1+s)n \sigma]
\mbox{,}
\eqno{(5.11)}
$$
where we have defined $\gamma_n(s)$ as

$$
\gamma_n^\mu(s) = \alpha_n^\mu(s) + \tilde{\alpha}_n^\mu(s), ~~~~n \neq 0
\mbox{.}
\eqno{(5.12)}
$$
The oscillations in region II are thus standing waves; this being a direct
consequence of the condition $x \rightarrow 0$.

It is useful to introduce light-cone coordinates, $\sigma^- = \tau - \sigma$ 
and $\sigma^+ = \tau + \sigma$. The derivatives conjugate to $\sigma^\mp$ are
$\partial_\mp = \frac{1}{2}(\partial_\tau \mp \partial_\sigma)$. In region I we
find

$$
\partial_{-}X^\mu = \frac{1+s}{2\sqrt{\pi T_I}} \sum_{-\infty}^{\infty}
\alpha_{n}^\mu(s)e^{i(1+s)n(\sigma-\tau)}
\eqno{(5.13)}
\mbox{,}
$$
$$
\partial_{+}X^\mu = \frac{1+s}{2\sqrt{\pi T_I}} \sum_{-\infty}^{\infty}
\tilde{\alpha_{n}}^\mu(s)e^{-i(1+s)n(\sigma+\tau)}
\mbox{,}
\eqno{(5.14)}
$$
where we have defined

$$
\alpha_0^{\mu}(s) = \tilde{\alpha}_0^\mu(s) = \frac{p^{\mu}}{T_{II}s}
\sqrt{\frac{T_I}{\pi}}
\mbox{.}
\eqno{(5.15)}
$$

Further, in region II  we find

$$
\partial_{\mp} X^\mu = \frac{1+s}{4 \sqrt{\pi T_I}}
\sum_{-\infty}^{\infty}
\gamma_n^\mu(s)
e^{\pm i(1+s) n (\sigma \mp \tau)}
\mbox{,}
\eqno{(5.16)}
$$
with

$$
\gamma_0^\mu(s) = \frac{2p^\mu}{T_{II}s} \sqrt{\frac{T_I}{\pi}} =
2 \alpha_0^\mu(s)
\mbox{.}
\eqno{(5.17)}
$$

Inserting Eqs. (5.12) and (5.16) into the Hamiltonian

$$
H = \int_0^{\pi} T(\sigma)(\partial_{-}X \cdot \partial_{-}X + 
\partial_{+}X \cdot \partial_{+}X) d\sigma
\eqno{(5.18)}
$$
we get, for the full first branch $H = H_I + H_{II}$, where

$$
H_I  =  T_I \int_{I} (\partial_-X \cdot \partial_-X + \partial_+X
\cdot \partial_+X) d\sigma
$$
$$
= \frac{1+s}{4} \sum_{-\infty}^{\infty} [\alpha_{-n}(s) \cdot \alpha_n(s)
+ \tilde{\alpha}_{-n}(s) \cdot \tilde{\alpha}_n(s)]
\mbox{,}
\eqno{(5.19)}
$$

$$
H_{II}  =  T_{II} \int_{II} (\partial_-X \cdot \partial_-X + \partial_+X
\cdot \partial_+X) d\sigma
$$
$$
= \frac{s(1+s)}{8x} \sum_{-\infty}^{\infty} \gamma_{-n}(s) \cdot
\gamma_n(s)
\mbox{,}
\eqno{(5.20)}
$$
with $x = T_I/T_{II}$ as before.

The case $s = 1$ is of particular interest. The string is then divided into
two pieces of equal length. We have then

$$
H_I(s=1) = \frac{1}{2}\sum_{-\infty}^{\infty} \left(
\alpha_{-n} \cdot \alpha_n + \tilde{\alpha}_{-n} \cdot \tilde{\alpha}_n \right)
\mbox{,}
\eqno{(5.21)}
$$

$$
H_{II}(s=1) = \frac{1}{4x} \sum_{-\infty}^{\infty} \gamma_{-n} \cdot \gamma_n
\mbox{.}
\eqno{(5.22)}
$$
It is notable that Eq. (5.21) is formally the same as the standard
expression for a closed uniform string, of length $\pi$. See, for instance,
Eq. (2.1.76) in Ref. \cite{green87}. The fact that we recover the
characteristics of a closed string in region I is understandable, since
this part of our composite string permits both right-moving and
left-moving waves. 
Eq. (5.22) is essentially the standard expression
for en {\it open} uniform string, corresponding to standing waves. The
presence of the inverse tension ratio $x^{-1}$ in front of the expression is
caused by the junction conditions, Eqs. (5.9).

The condition $H=0$ enables us to calculate the mass $M$ of the string. It
must be given by $M^2 = -p^\mu p_\mu$, similarly as in the uniform string case
\cite{green87}. From Eqs. (5.19) and (5.20) we obtain, taking into
account that $x << 1$ and that
$\alpha_0(s) \cdot \alpha_0(s) =-M^2x/(\pi T_{II} s^2)$,

$$
M^2 = \pi T_{II} s  \sum_{n=1}^{\infty} \left[ \alpha_{-n}(s) \cdot
\alpha_n(s) + \tilde{\alpha}_{-n}(s) \cdot \tilde\alpha_n(s) +
\frac{s}{2x} \gamma_{-n}(s) \cdot \gamma_n(s) \right]
\mbox{.}
\eqno{(5.23)}
$$
This holds for the first branch, for odd/even values of $s$.

\subsection{The second branch}

For the second branch whose eigenfrequencies are $\omega (s^{-1})$ the mode 
expansion in region I becomes 

$$
X_I^\mu = \frac{i}{2\sqrt{\pi T_I}}\sum_{n \neq 0} \frac{1}{n}
\left[
\alpha_{n}^\mu(s^{-1})e^{i(1+s^{-1})n(\sigma-\tau)} +
\tilde{\alpha}_{n}^\mu(s^{-1})e^{-i(1+s^{-1})n(\sigma+\tau)}
\right]
\mbox{.}
\eqno{(5.24)}
$$

Analogously in region II

$$
X_{II}^\mu = \frac{i}{2\sqrt{\pi T_I}} \sum_{n \neq 0} \frac{1}{n}
\gamma_{n}^\mu(s^{-1})e^{-i(1+s^{-1})n\tau} \cos (1+s^{-1})n\sigma
\mbox{,}
\eqno{(5.25)}
$$

where

$$
\gamma_{n}^\mu(s^{-1}) = \alpha_{n}^\mu(s^{-1}) + \tilde{\alpha}_{n}^
\mu(s^{-1}),~~ n\neq 0
\mbox{.}
\eqno{(5.26)}
$$
The expansions (5.24) and (5.25) hold for all integers $s$. This is so because
the basic
expressions (4.8) and (4.9) for the eigenfrequencies hold for all values of $s$.
However it may be noted that if the junction conditions are required to imply
nonvanishing oscillations in region II, corresponding to nonvanishing right
hand sides
in Eq. (5.9), then further restrictions come into play. Namely, if $s$ is odd,
the index $n$ in Eqs. (5.24) and (5.25) has to be a multiple of $s$.
If $s$ is even, then $n$ has to be an {\it even} integer times $s$.
We recall that analogous considerations were made in the case of the first
branch.
When we later shall consider the quantum mechanical free energy, it becomes
necessary
to include {\it all} modes, including those that lead to zero oscillations in
region II
according to the classical theory.

Let us calculate the light-cone derivatives: in region I they are

$$
\partial_{-}X^\mu = \frac{1+s^{-1}}{2\sqrt{\pi T_I}} \sum_{-\infty}^{\infty}
\alpha_{n}^{\mu}(s^{-1})e^{i(1+s^{-1})n(\sigma-\tau)}
\mbox{,}
\eqno{(5.27)}
$$
$$
\partial_{+}X^\mu =  \frac{1+s^{-1}}{2\sqrt{\pi T_I}} \sum_{-\infty}^{\infty}
\tilde{\alpha_{n}^{\mu}}(s^{-1})e^{-i(1+s^{-1})n(\sigma+\tau)}
\mbox{,}
\eqno{(5.28)}
$$
and in region $II$

$$
\partial_{\mp}X^\mu = \frac{1+s^{-1}}{4\sqrt{\pi T_I}} \sum_{-\infty}^{\infty}
\gamma_{n}^\mu(s^{-1})e^{\pm i(1+s^{-1})n(\sigma \mp \tau)}
\mbox{,}
\eqno{(5.29)}
$$
where

$$
\alpha_0^\mu(s^{-1}) = \tilde{\alpha_0^\mu}(s^{-1})= \frac{1}{2} \gamma_0^
\mu(s^{-1}) = \frac{p^\mu}{T_{II}} \sqrt{\frac{T_I}{\pi }}
\mbox{.}
\eqno{(5.30)}
$$
Thus $\alpha_0(s^{-1})$ differs from $\alpha_0(s)$, Eq. (5.15).
Again writing the Hamiltonian as $H = H_I + H_{II}$, we now find

$$
H_I = \frac{1+s^{-1}}{4s} \sum_{-\infty}^{\infty}
\left[
\alpha_{-n}(s^{-1}) \cdot \alpha_{n}(s^{-1}) + \tilde{\alpha}_{-n}(s^{-1}) 
\cdot\tilde{\alpha}_{n}(s^{-1})\right]
\mbox{,}
\eqno{(5.31)}
$$
$$
H_{II} = \frac{1+s^{-1}}{8x} \sum_{-\infty}^{\infty}
\gamma_{-n}(s^{-1}) \cdot \gamma_{n}(s^{-1})
\mbox{.}
\eqno{(5.32)}
$$
If $s=1$, we recover the same expressions for $H_I$ and $H_{II}$, Eqs.
(5.21) and (5.22), as for the first branch.

From the condition $H=0$ we calculate the mass, observing that
$\alpha_0(s^{-1}) \cdot \alpha_0(s^{-1}) = -M^2x/(\pi T_{II}$):

$$
M^2 =  \frac{\pi T_{II}}{s} \sum_{n=1}^{\infty}
\left[
\alpha_{-n}(s^{-1}) \cdot \alpha_{n}(s^{-1}) + \tilde{\alpha}_{-n}(s^{-1})\cdot
\tilde{\alpha}_{n}(s^{-1})\right] 
$$
$$
+ \frac{\pi T_{II}}{2x}\sum_{n=1}^{\infty}\gamma_{-n}(s^{-1})\cdot
\gamma_{n}(s^{-1})
\mbox{.}
\eqno{(5.33)}
$$

\section{Quantum theory. The free energy of the string}

\subsection{Quantization}

We shall consider the free energy of the quantum fields with masses given
by the mass formula corresponding to the piecewice bosonic string. 
We quantize the system according to conventional methods as
found, for instance, in Ref. \cite{green87},
Chapter 2.2. In accordance with the canonical prescription in region I the 
equal-time commutation rules are required to be

$$
T_{I}[\dot{X}^\mu(\sigma,\tau), X^\nu(\sigma', \tau)]=
-i \delta(\sigma-\sigma')\eta^{\mu \nu}
\mbox{,}
\eqno{(6.1)}
$$
and in region II
$$
T_{II} [\dot{X}^\mu(\sigma,\tau),X^\nu(\sigma',\tau)] =
-i \delta(\sigma - \sigma')\eta^{\mu \nu}
\mbox{,}
\eqno{(6.2)}
$$
where $\eta^{\mu \nu}$ is the $D$-dimensional metric. These relations are in
conformity with the fact that the momentum conjugate to $X^\mu$ is in either
region equal to $T(\sigma) \dot{X}^\mu$. The remaining commutation relations
vanish:

$$
[X^\mu(\sigma,\tau),X^\nu(\sigma',\tau)] = [\dot{X}^\mu(\sigma,\tau),
\dot{X}^\nu(\sigma',\tau)]= 0
\mbox{.}
\eqno{(6.3)}
$$

The quantities to be promoted to Fock state operators are
$\alpha_{\mp n}(s)$ and $\tilde{\alpha}_{\mp n}(s)$ (first branch, region I),
$\gamma_{\mp n}(s)$ (first branch, region II),
$\alpha_{\mp n}(s^{-1})$ and $\tilde{\alpha}_{\mp n}(s^{-1})$ (second branch,
region I), and $\gamma_{\mp n}(s^{-1})$ (second branch, region II). These
operators satisfy

$$
\alpha_{-n}^{\mu}(s) = \alpha_{n}^{\mu\dagger}(s),\,\,\,
\gamma_{-n}^{\mu}(s) = \gamma_{n}^{\mu\dagger}(s)
\mbox{,}
\eqno{(6.4)}
$$
$$
\alpha_{-n}^{\mu}(s^{-1}) = \alpha_{n}^{\mu\dagger}(s^{-1}),\,\,\,
\gamma_{-n}^{\mu}(s^{-1}) = \gamma_{n}^{\mu\dagger}(s^{-1})
\eqno{(6.5)}
$$
for all $n$.
We insert our previous expansions for  $X^\mu$ and $\dot{X}^\mu$ in the
commutation relations in regions I and II for the two branches, and make use
of the effective relationship

$$
\sum_{-\infty}^{\infty}e^{i(1+s)n(\sigma-\sigma')}=
2\sum_{-\infty}^{\infty}\cos(1+s)n\sigma\cos(1+s)n\sigma'
\rightarrow \frac{2\pi}{1+s}\delta(\sigma-\sigma')
\mbox{.}
\eqno{(6.6)}
$$
For the first branch we then get in region I

$$
[\alpha_n^\mu(s),\alpha_m^\nu(s)] = n \delta_{n+m,0} \eta^{\mu\nu}
\mbox{,}
\eqno{(6.7)}
$$
with a similar relation for the $\tilde{\alpha}_n$. In region II

$$
[\gamma_{n}^\mu(s),\gamma_{m}^\nu(s)] = 4nx \delta_{n+m,0} \eta^{\mu\nu}
\mbox{.}
\eqno{(6.8)}
$$
For the second branch we get analogously

$$
[\alpha_{n}^{\mu}(s^{-1}),\alpha_{m}^{\nu}(s^{-1})]=n\delta_{n+m,0}
\eta^{\mu\nu},\,\,\,\,\,
[\gamma_{n}^{\mu}(s^{-1}), \gamma_{m}^{\nu}(s^{-1})]= 
4nx\delta_{n+m,0}\eta^{\mu\nu}
\mbox{.}
\eqno{(6.9)}
$$
By introducing annihilation and creation operators for the first branch in the
following way:

$$
\alpha_n^\mu(s) = \sqrt{n} a_n^\mu(s),\,\,\,\,\,
\alpha_{-n}^\mu(s)=\sqrt{n} a_n^{\mu\dagger}(s)
\mbox{,}
\eqno{(6.10)}
$$
$$
\gamma_n^\mu(s) = \sqrt{4nx} c_{n}^{mu}(s),\,\,\,\,\,
\gamma_{-n}^\mu(s) =
\sqrt{4nx} c_{n}^{\mu\dagger}(s)
\mbox{,}
\eqno{(6.11)}
$$
we find for $n \geq 1$ the standard form

$$
[a_{n}^{\mu}(s), a_{m}^{\nu\dagger}(s)] = \delta_{nm}\eta^{\mu\nu}
\eqno{(6.12)}
$$

$$
[c_{n}^{\mu}(s), c_{m}^{\nu\dagger}(s)] = \delta_{nm}\eta^{\mu\nu}
\mbox{.}
\eqno{(6.13)}
$$
The commutation relations for the second branch are analogous,
only with the replacement $s\rightarrow s^{-1}$.

\subsection{The free energy and the Hagedorn temperature}

In the following we shall limit ourselves to the first branch only. Using Eqs.
(6.10) and (6.11) in
Eqs.(5.19) and (5.20) we may write the two parts of the Hamiltonian as

$$
H_{I} = -\frac{M^2x}{2st(s)} + \frac{1}{2}\sum_{n=1}^{\infty} \omega_n(s)
[a_n^{\dagger}(s)\cdot a_n(s) + \tilde{a_n}^{\dagger}(s)\cdot\tilde{a_n}(s)]
\mbox{,}
\eqno{(6.14)}
$$

$$
H_{II} = -\frac{M^2}{2t(s)} + s \sum_{n=1}^{\infty} \omega_n(s)
c_n^{\dagger}(s) \cdot c_n(s)
\mbox{,}
\eqno{(6.15)}
$$
where we for convenience have introduced the symbol $t(s)$ defined by
$t(s) = \pi\bar{T}(s)$.
(Observe the notation  $c_n^{\dagger}\cdot c_n \equiv c_n^{\mu\dagger}
c_{n\mu}$). The extra factor $s$ in Eq. (6.15) is
related to the fact that the relative length of region II is equal to
$s$.
From the condition $H = H_{I}+H_{II} = 0$ in the limit $x \rightarrow 0$
we obtain, either from Eqs. (6.14) and (6.15) or directly from Eq. (5.23),

$$
M^{2} = t(s)\sum_{i=1}^{24}\sum_{n=1}^{\infty}\omega_n(s)
[a_{ni}^{\dagger}(s)a_{ni}(s) + \tilde{a}_{ni}^{\dagger}\tilde{a}_{ni}(s)
- A_1] 
$$
$$
+ 2st(s)\sum_{i=1}^{24}\sum_{n=1}^{\infty}\omega_n(s)[c_{ni}^{\dagger}(s)
c_{ni}(s)- A_2]
\mbox{.}
\eqno{(6.16)}
$$
We have here put $D = 26$, the commonly accepted space-time dimension for the
bosonic string.
As usual, the $c_{ni}$ denote the transverse oscillator operators  (here for
the first branch). Further, we have introduced in Eq. (6.16) two constants
$A_1$ and $A_2$, in order to take care of ordering ambiguities.

A zero-point energy $\frac{1}{2} \sum \omega_n$,
summed over all eigenfrequencies, is actually the Casimir energy, which was
calculated in \cite{brevniels90}. When $x \rightarrow 0$ we have, for
arbitrary $s$, when the string length equals $\pi$,

$$
\frac{1}{2} \sum_{- \infty}^{\infty} \omega_n \rightarrow
- \frac{1}{24}(s + \frac{1}{s} - 2)
\mbox{.}
\eqno{(6.17)}
$$

The constraint for the closed string (expressing the invariance of the theory
in the region I under shifts of the origin of the co-ordinate) has the form

$$
\sum_{i=1}^{24}\sum_{n=1}^{\infty}\omega_n(s)\left[
a_{ni}^{\dagger}(s)a_{ni}(s) - \tilde{a}_{ni}^{\dagger}\tilde{a}_{ni}(s)
\right]=0
\mbox{.}
\eqno{(6.18)}
$$
The commutation relations for above operators are given by Eqs. (6.12) and 
(6.13).
The mass of state (obtained by acting on the Fock vacuum $|0>$ with creation
operators) can be written as follows
$({\rm mass})^2\sim a_{n1}^{\dagger}...a_{ni}^{\dagger}c_{n1}^{\dagger}...
c_{ni}^{\dagger}|0>$.

Let us start with the discussion of the free energy in field theory
at non-zero temperature. It is quite well-known that the one-loop
free energy for the bosonic (b) or fermionic (f) degree of freedom
in d-dimensional space is given by

$$
{\frak F}_{b,f} = \pm \frac{1}{\beta} \int \frac{d^{d-1} k}{(2\pi )^{d-1}}
\log \left(1\mp  e^{-\beta u_k} \right)
\mbox{,}
\eqno{(6.19)}
$$
where $\beta=(k_BT)^{-1}$,\, $u_k=\sqrt{k^2+m^2}$, and $m$ is the mass for 
the corresponding degree of freedom.
Expanding the logarithm and performing the (elementary) integration
one easily gets 

$$
{\frak F}_b = - \sum_{n=1}^{\infty} (\beta n)^{-d/2} \pi^{-d/2} 2^{1-d/2}
m^{d/2} K_{d/2} (\beta n m)
\mbox{,}
\eqno{(6.20)}
$$
$$
{\frak F}_f = - \sum_{n=1}^{\infty} (-1)^n (\beta n)^{-d/2} \pi^{-d/2}
2^{1-d/2} m^{d/2} K_{d/2} (\beta n m)
\mbox{,}
\eqno{(6.21)}
$$
where $ K_{d/2} (z)$ are the modified Bessel functions. Using the
integral representation for the Bessel function

$$
K_{d/2} (z) = \frac{1}{2} \left( \frac{z}{2} \right)^{d/2}
\int_0^{\infty} ds \, s^{-1-d/2} e^{-s-z^2/(4s)}
\mbox{,}
\eqno{(6.22)}
$$
one can obtain the well-known proper time representation for the
one-loop free energy:

$$
{\frak F}_b = - \int_0^{\infty} ds \, \pi^{-d/2} 2^{-1-d/2}  s^{-1-d/2}
e^{-m^2s/2} \left[ \theta_3  \left( 0 |
\frac{i\beta^2}{2\pi s} \right) -1 \right]
\mbox{,}
\eqno{(6.23)}
$$
$$
{\frak F}_f = - \int_0^{\infty} ds \, \pi^{-d/2} 2^{-1-d/2}  s^{-1-d/2}
e^{-m^2s/2} \left[1- \theta_4  \left( 0 |
\frac{i\beta^2}{2\pi s} \right)\right]
\mbox{,}
\eqno{(6.24)}
$$
where $\theta_3(v|x) = \sum_{n=-\infty}^{\infty}\exp\left(ixn^2+2 \pi ivn
\right)$ and $\theta_4(v|x)=\theta_3(v+1/2|x)$ are the Jacobi theta
functions.
Expressions (6.23) and (6.24) is usually the starting point for the calculation
of the (super) string free energy in the canonical ensemble (then $m^2$ is
the mass operator and for closed strings the corresponding
constraint  should be taken into account).

As usual the physical Hilbert space consists of all Fock
space states obeying the condition (6.18), which can be implemented by means
of the integral representation for Kronecker deltas. Thus the free energy
of the field content in the "proper time" representation becomes

$$
F =-\frac{1}{24}(s+\frac{1}{s}-2)
$$
$$ 
-2^{-14}\pi^{-13}\int_0^{\infty}\frac{d\tau_2}{\tau_2^{14}}
\left[\theta_3\left(0|\frac{i\beta^2}{2\pi\tau_2}\right)-1\right]
{\rm Tr}\exp\left\{-\frac{\tau_2M^2}{2}\right\}
$$
$$
\times \int_{-\pi}^{\pi}\frac{d\tau_1}{2\pi}{\rm Tr}
\exp\left\{i\tau_1\sum_{i=1}^{24}\sum_{n=1}^{\infty}\omega_n(s)\left[
a_{ni}^{\dagger}(s)a_{ni}(s) - \tilde{a}_{ni}^{\dagger}(s)\tilde{a}_{ni}(s)
\right]\right\}
\mbox{.}
\eqno{(6.25)}
$$
Performing the trace over the entire Fock space (note that $[H_I,H_{II}]=0$
and ${\rm Tr}\,y^{a_n^{\dagger}a_n}=(1-y)^{-1}$) we have

$$
{\rm Tr}\exp\left\{\sum_{i=1}^{24}\sum_{n=1}^{\infty}\omega_n(s)
a_{ni}^{\dagger}(s)a_{ni}(s)\left(-\frac{1}{2}t(s)\tau_2\pm i\tau_1\right)
\right\}
$$
$$
=\prod_{n=1}^{\infty}\left[1-e^{\omega_n(s)(-\frac{1}{2}t(s)\tau_2\pm i\tau_1)}
\right]^{-24}
\mbox{,}
\eqno{(6.26)}
$$
$$
{\rm Tr}\exp\left\{-st(s)\tau_2\sum_{i=1}^{24}\sum_{n=1}^{\infty}\omega_n(s)
c_{ni}^{\dagger}(s)c_{ni}(s)\right\}
$$
$$
=\prod_{n=1}^{\infty}\left[1-e^{-st(s)\tau_2\omega_n(s)}\right]^{-24}
\mbox{.}
\eqno{(6.27)}
$$
Working out the sums in Eq. (6.25) for $A_1=2,\,\,A_2=1$, and changing 
variables to $\tau_1\rightarrow\tau_1 2\pi,\,\tau_2\rightarrow 
\tau_2 4\pi/t(s)$ one can finally get

$$
F=-\frac{1}{24}(s+\frac{1}{s}-2) 
-2^{-40}\pi^{-26}t(s)^{-13}\int_0^{\infty}\frac{d\tau_2}{\tau_2^{14}}
\int_{-1/2}^{1/2}d\tau_1
$$
$$
\times\left[\theta_3\left(0|\frac{i\beta^2t(s)}
{8\pi^2\tau_2}\right)-1\right]|\eta[(1+s)\tau]|^{-48}\eta[s(1+s)
(\tau-\overline{\tau})]^{-24}
\mbox{,}
\eqno{(6.28)}
$$
where we integrate over all possible non-diffeomorphic toruses which are
characterized by a single Teichm{\"u}ller parameter $\tau=\tau_1+i\tau_2$.
In Eq. (6.28) the condition $\eta(-\overline{\tau})=\overline{\eta(\tau)}$ 
has been used.

Once the free energy has been found, the other thermodynamic quantities can
readily be calculated. For instance, the energy $U$ and the entropy $S$ of
the system are

$$
U = \frac{\partial (\beta F)}{\partial \beta},
~~~~~~~~ S = k_B \beta^2 \frac{\partial F}{\partial \beta}
\mbox{.}
\eqno{(6.29)}
$$

What is the Hagedorn temperature, $T_c = 1/(k_B \beta_c)$, of the composite
string? This critical temperature, introduced by Hagedorn in the context
of strong interactions a long time ago \cite{hagedorn65}, is the
temperature above which the free energy is ultraviolet divergent. In the
ultraviolet limit ($\tau_2 \rightarrow 0$),

$$
\eta^{-24} (i \tau) = \tau^{12} e^{2 \pi/ \tau}\left[1+O\left
(e^{-2\pi/\tau}\right)\right]
\mbox{,}
\eqno{(6.30)}
$$
$$
\theta_3 \left( 0|\frac{i\beta^2t(s)}{8\pi^2\tau_2} \right)-1
= 2 \exp \left( - \frac{\beta^2t(s)}{8\pi^2\tau_2}\right)+
O\left(\exp\left(-\frac{\beta^2t(s)}{2\pi^2\tau_2}\right)\right)
\mbox{,}
\eqno{(6.31)}
$$
which upon insertion into Eq. (6.28) shows that the integrand is ultraviolet
finite if

$$
\beta > \beta_c = \frac{4}{s}\sqrt{\frac{\pi(1+s)}{T_{II}}}
\mbox{.}
\eqno{(6.32)}
$$

For a fixed value of $T_{II}$ the Hagedorn temperature is thus seen to depend
on $s$. We may mention here that the physical meaning of the Hagedorn
temperature is still not clear. There are different interpretations
possible: {\it (i)} one may argue that $T_c$ is the maximum obtainable
temperature in string systems, this meaning, when applied to cosmology,
that there is a maximum temperature in the early Universe. Or, {\it (ii)} one
may take $T_c$ to indicate some sort of phase transition to a new stringy
phase. Some further discussion on these matters is given, for instance,
in Refs. \cite{alvarez87,elidrom,bytsenko96}.

Finally, let us consider the limiting case in which one of the pieces of the
string is much shorter than the other. Physically this case is of interest,
since it corresponds to a point mass sitting on a string. Since we have
assumed that $s \geq 1$, this case corresponds to $s \rightarrow\infty$. We
let the tension $T_{II}$ be fixed, though arbitrary.
It is seen, first of all, that the Hagedorn temperature (6.32) goes to
infinity so that $F$ is always ultraviolet finite,
$\beta_c \rightarrow 0, ~~~~~~~~ T_c \rightarrow \infty$.
Next, since $\exp\left(-\beta^2t(s)/8\pi^2\tau_2\right)$ can be
taken to be small we obtain, when using again the expansion (6.31) for
$\theta_3\left(0|i\beta^2t(s)/8\pi^2\tau_2\right)$,

$$
F_{(\beta\rightarrow 0)} = -\frac{s}{24}-
(8\pi^3T_{II})^{-13}\int_0^{\infty}\frac{d\tau_2}{\tau_2^{14}}
\int_{-1/2}^{1/2}d\tau_1 
$$
$$
\times  \exp\left(-\frac{\beta^2T_{II}}{8\pi\tau_2}\right)
|\eta[(1+s)\tau]|^{-48}\eta[s(1+s)(\tau-\overline{\tau})]^{-24}
\mbox{.}
\eqno{(6.33)}
$$
Physically speaking, the linear dependence of the first term in (6.33)
reflects that the Casimir energy of a little piece of string
embedded in an essentially infinite string has for dimensional reasons to be
inversely proportional to the length $L_I = \pi/(1+s) \simeq \pi/s$ of
the little string. The first term in (6.33) is seen to outweigh the
second, integral term, which goes to zero when $s \rightarrow \infty$.


\newpage
\begin{thebibliography}{99}

\bibitem{brevniels90}
{\sc I. Brevik and H. B. Nielsen}, Phys. Rev. D {\bf 41}, 1185 (1990).

\bibitem{li91}
{\sc X. Li, X. Shi and J. Zhang}, Phys. Rev. D {\bf 44}, 560 (1991).

\bibitem{brevik94}
{\sc I. Brevik and E. Elizalde}, Phys. Rev. D {\bf 49}, 5319 (1994).

\bibitem{brevik95}
{\sc I. Brevik and H.B. Nielsen}, Phys. Rev. D {\bf 51}, 1869 (1995).

\bibitem{brevik96}
{\sc I. Brevik, H.B. Nielsen and S.D. Odintsov}, Phys. Rev. D {\bf 53}, 3224 
(1996).

\bibitem{bayin96}
{\sc S.S. Bayin, J.P. Krisch and M. Ozcan},
J. Math. Phys. {\bf 37}, 3662 (1996).

\bibitem{brevik}
{\sc I. Brevik and R. Sollie}, J. Math. Phys. {\bf 38}, 2774 (1997).

\bibitem{berntsen97}
{\sc M.H. Berntsen, I. Brevik and S.D. Odintsov}, Ann. Phys. (NY) {\bf 257}, 
84 (1997).

\bibitem{lu}
{\sc J. Lu and B. Huang}, Phys. Rev. D  {\bf 57}, 5280 (1998).

\bibitem{brevik98}
{\sc I. Brevik, A. A. Bytsenko, and H. B. Nielsen}, Class. Quant. Grav.
{\bf 15}, 1 (1998).

\bibitem{brevclaus}
{\sc I. Brevik and I. Clausen}, Phys. Rev. D  {\bf 39}, 603 (1989).

\bibitem{vankampen}
{\sc N. G. van Kampen, B. R. A. Nijboer, and K. Schram}, Phys. Lett. A 
{\bf 26}, 307 (1968).

\bibitem{elidrom}
{\sc E. Elizalde, S. D. Odintsov, A. Romeo, A. A. Bytsenko, and S. Zerbini}, 
{\em ``Zeta Regularization Techniques with Applications''} (World Scientific,  
Singapore, 1994).

\bibitem{bytsenko96}
{\sc A.A. Bytsenko, G. Cognola, L. Vanzo and S. Zerbini},
Phys. Reports {\bf 266}, 1 (1996).

\bibitem{green87}
{\sc M.B. Green, J.H. Schwarz and E. Witten},
{\em ``Superstring Theory''}, Vol. {\bf 1} (Cambridge Univ. Press, Cambridge, 
1987).

\bibitem{hagedorn65}
{\sc R. Hagedorn}, Suppl. Il Nuovo Cimento {\bf 3}, 147 (1965).

\bibitem{alvarez87}
{\sc E. Alvarez and M.A.R. Osorio}, Phys. Rev. D {\bf 36}, 1175 (1987).


\end{thebibliography}
\end{document}